\documentclass[12pt,showkeys,showpacs,indentfirst,bm]{revtex4}
\usepackage{epsfig}
\usepackage{epsf}
\usepackage{multirow}

\begin{document}

\title{Dynamically generated $1^+$ heavy mesons}

\author{Feng-Kun Guo$^{1,2,6}$}
\email{guofk@mail.ihep.ac.cn}
\author{Peng-Nian Shen$^{2,1,4,5}$}
\author{Huan-Ching Chiang$^{3,1}$}
\affiliation{\small $^1$Institute of High Energy Physics, Chinese
Academy of Sciences,
P.O.Box 918(4), Beijing 100049, China\footnote{Corresponding address.}\\
$^2$CCAST(World Lab.), P.O.Box 8730, Beijing 100080, China\\
$^3$South-west University, Chongqing 400715, China\\
$^4$Institute of Theoretical Physics, Chinese Academy of Sciences, P.O.Box 2735, China\\
$^5$Center of Theoretical Nuclear Physics, National Laboratory of
Heavy Ion Accelerator, Lanzhou 730000, China\\
$^6$Graduate University of Chinese Academy of Sciences, Beijing
100049, China}
\date{\today}

\begin{abstract}
By using a heavy chiral unitary approach, we study the $S$ wave
interactions between heavy vector meson and light pseudoscalar
meson. By searching for poles of the unitary scattering amplitudes
in the appropriate Riemann sheets , several $1^+$ heavy states are
found. In particular, a $D^*K$ bound state with a mass of
$2.462\pm0.010$ GeV which should be associated with the recently
observed $D_{s1}(2460)$ state is obtained. In the same way, a
$B^*{\bar K}$ bound state ($B_{s1}$) with mass of $5.778\pm0.007$
GeV in the bottom sector is predicted. The spectra of the
dynamically generated $D_1$ and $B_1$ states in the $I=1/2$ channel
are also calculated. One broad state and one narrow state are found
in both the charmed and bottom sectors. The coupling constants and
decay widths of the predicted states are further investigated.

\end{abstract}


\maketitle

\section{Introduction}

In recent years, the discovery of various new hadron states
stimulated much theoretical effort on the hadron spectrum. Among
these states, $D_{s0}^*(2317)$ and $D_{s1}(2460)$
\cite{prl90} 
are the most attractive ones, because the measured masses are
smaller than those predicted in terms of most phenomenological
models \cite{go03}(refer to the recent review articles \cite{sw06}).
Many physicists presumed that these new states are conventional
$c{\bar s}$ mesons
\cite{go03,be03}, 
and the others believed that they might be exotic meson states, such
as tetraquark states \cite{ch03}, 
molecular states \cite{sz03,kl04,hl04,ro06}, or admixture of $c{\bar
s}$ with molecular component or tetraquark component \cite{ww63}, 
and etc..

In the corresponding non-strange sector, there are two confirmed
$D_1$ states. The narrow one is named as $D_1(2420)$ with mass of
$2423.4\pm3.1$ MeV and width of $25\pm6$ MeV, and the broad one is
entitled as $D_1(2430)$ with mass of $2427\pm26\pm25$ MeV and width
of $384^{+107}_{-75}\pm74$ MeV \cite{be04,pdg06}. Theoretically, the
wide-width state with mass of $2325$ MeV and narrow-width state with
mass of $2552$ MeV were declared in the $\chi$-SU(3) approach
\cite{kl04}. In Ref. \cite{hl04}, these two states were considered
as quasi-bound states and were used to determine the unknown
coupling constants in the next-to-leading order heavy chiral
Lagrangian. The reproduced masses and widths are $M_{D_1}=2422$ MeV
and $\Gamma_{D_1}=23$ MeV and $M_{D_1'}=2300$ MeV and
$\Gamma_{D_1'}=300$ MeV, respectively. In many other references
\cite{gi85,cl94,pe01,go05,cs05}, these two states were proposed as
the conventional $c{\bar q}$ states, for instance, in Ref.
\cite{cs05}, they were deliberated as the mixed $^1P_1$ and $^3P_1$
$c{\bar q}$ states with a mixing angle of $\phi\approx 35^{\circ}$
obtained by fitting measured widths.

Since the $D_{s0}^*(2317)$ and $D_{s1}^*(2460)$ have the same
quantum number $I$, the isospin, and $S$, the strangeness, except
$J$, the total angular momentum, in the $S$ wave $DK$ and $D^*K$
channels, respectively, it would be constructive to study the $DK$
and $D^*K$ interactions so that one can see whether these two states
could have similar molecular structure. In the light hadron system,
the chiral unitary approach (ChUA) has achieved great success in
explaining the meson-meson and meson-baryon interactions
\cite{kw97,oo97,oo99,oond,ka98,mo00,ro05,gp05}. Some well known
hadrons can be dynamically generated as the quasi-bound states of
two mesons or a meson and a baryon \cite{ro05}, for instance, the
lowest scalar states $\sigma$, $f_0(980)$, $a_0(980)$, $\kappa$
\cite{oo97,oo99,oond,gp05}, and etc.. Then, the approach was
extended to the heavy hadron system, called heavy chiral unitary
approach (HChUA) \cite{gs06}. In terms of such an approach, the $S$
wave interaction between the heavy meson and light pseudoscalar
meson was studied, and some bound states and resonances were
predicted, for example, the $D_{s0}^*(2317)$ state as a $DK$ bound
state at $2.312\pm0.041$ GeV, a $B{\bar K}$ bound state at
$5.725\pm0.039$ GeV \cite{gs06}. In the same approach, $D_0^*$ and
$B_0^*$ in the $(I,~S)=(1/2,~0)$ channel were also investigated. As
a result, one broad state and one narrow state were predicted in
both the charmed and bottom sectors \cite{gs06}.

In this paper, we study the $S$ wave interaction between heavy
vector meson and light pseudoscalar meson and search for $J^P=1^+$
heavy mesons in both strange and non-strange sectors. The couplings
of various coupled channels to the generated states and the decay
width of the isospin symmetry violating process $D_{s1}(2460)^+\to
D_s^{*+}\pi^0$ are studied as well.

\section{Dynamically generated heavy-light $1^+$ states}

To describe the interaction between the heavy vector meson and the
light pseudoscalar meson, we employ a leading order heavy chiral
Lagrangian \cite{bd92,kl04}
\begin{equation}
\label{eq:L} {\cal L} =
-\frac{1}{4f_{\pi}^2}(\partial^{\mu}P^{*\nu}[\Phi,\partial_{\mu}\Phi]P^{*\dag}_{\nu}
- P^{*\nu}[\Phi,\partial_{\mu}\Phi]\partial^{\mu}P^{*\dag}_{\nu}),
\end{equation}
where $f_{\pi}=92.4$ MeV is the pion decay constant, $P^*_{\nu}$
represents the heavy vector mesons with quark contents $(Q{\bar
u},~Q{\bar d},~Q{\bar s})$, namely $(D^{*0},~D^{*+},~D_s^{*+})$ and
$(B^{*-},~{\bar B}^{*0},~{\bar B}_s^*)$ for the charmed and bottom
sectors, respectively, and $\Phi$ denotes the octet Goldstone bosons
in the $3\times3$ matrix form
\begin{equation}
\label{eq:ps} \Phi = \left(
\begin{array}{ccc}
\frac{1}{\sqrt{2}}\pi^0 + \frac{1}{\sqrt{6}}\eta & \pi^+ & K^+\\
\pi^- & -\frac{1}{\sqrt{2}}\pi^0 + \frac{1}{\sqrt{6}}\eta & K^0\\
K^- & \bar{K}^0 & - \frac{2}{\sqrt{6}}\eta
\end{array}
\right).
\end{equation}

We are interested in $(I,~S)=(0,~1)$ and $(I,~S)=(1/2,~0)$ systems.
Usually, such a system can be characterized by its own isospin.
Based on the Lagrangian in Eq. (\ref{eq:L}), the amplitude with a
definite isospin can be written as
\begin{equation}
\label{eq:amp} V^{I}_{ij}(s,t,u)\varepsilon\cdot\varepsilon' = -
\frac{C^{I}_{ij}}{4f_{\pi}^2}(s-u)\varepsilon\cdot\varepsilon',
\end{equation}
where $i$ and $j$ denote the initial state and the final state,
respectively, $s,~t,~u$ are usual Mandelstam variables, and
$\varepsilon$ and $\varepsilon'$ are polarization vectors of the
vector states in the initial and final states, respectively. In the
$I=0$ case, there are two coupled channels. The channel label $i=1$
and $2$ specify the $D^*K$ ($B^*{\bar K}$) and $D_s^*\eta$
($B_s^*\eta$) channels in the charmed (bottom) sector, respectively.
In the $I=1/2$ case, three coupled channels exist. The channel label
$i=1$, $2$ and $3$ in this case denote the $D^*\pi$ ($B^*\pi$),
$D^*\eta$ ($B^*\eta$) and $D_s^*{\bar K}$ ($B_s^*K$) in the charmed
(bottom) sector, respectively. The corresponding coefficients
$C^{I}_{ij}$ are tabulated in Table \ref{tab:cij}. It should be
mentioned that, in the coupled channel calculation, the thresholds
of the channels with the light vector meson and the heavy
pseudoscalar one are relatively higher than those with the light
pseudoscalar meson and the heavy vector one. For instance, in the
charmed $I=1/2$ case, the lightest combination of a light vector
meson and heavy pseudoscalar meson is $\rho+D$, and the sum of their
masses are almost forty MeV heavier than that of the heaviest
combination of a light pseudoscalar meson and heavy vector meson,
say $K+D_s^*$. Thus, in the concerned energy region near the
thresholds of the channels with later combinations, the
contributions from the channels with former combinations are
expected to be less important, and hence can be neglected for
simplicity.
\begin{table}[hbt]
\caption{\label{tab:cij} Coefficients $C^{I}_{ij}$ in
Eq.~(\ref{eq:amp}).}
\begin{center}
\begin{tabular}{ccc|cccccc}
\hline\hline $C^{0}_{11}$ & $C^{0}_{12}$ & $C^{0}_{22}$ &
$C^{1/2}_{11}$ & $C^{1/2}_{12}$
 & $C^{1/2}_{22}$ & $C^{1/2}_{13}$
 & $C^{1/2}_{23}$ & $C^{1/2}_{33}$\\
\hline $-2$ & $\sqrt{3}$ & 0 & $-2$ & 0 & 0 & $-\frac{\sqrt{6}}{2}$
& $-\frac{\sqrt{6}}{2}$ & $-1$\\
\hline\hline
\end{tabular}
\end{center}
\end{table}

\begin{figure}[hbt]
\begin{center}\vspace*{0.5cm}
{\epsfysize=4.5cm \epsffile{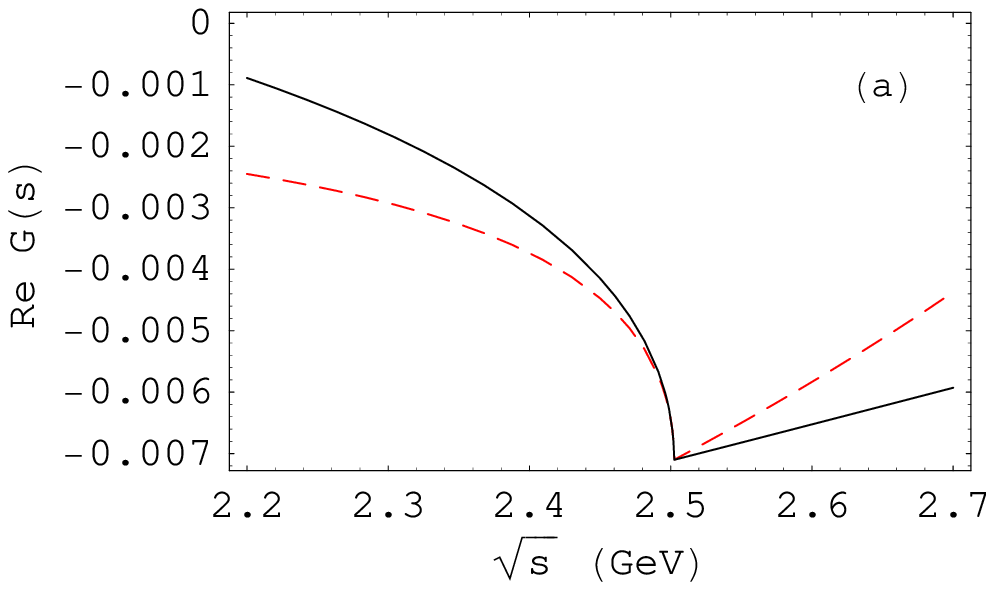}}%
{\epsfysize=4.5cm \epsffile{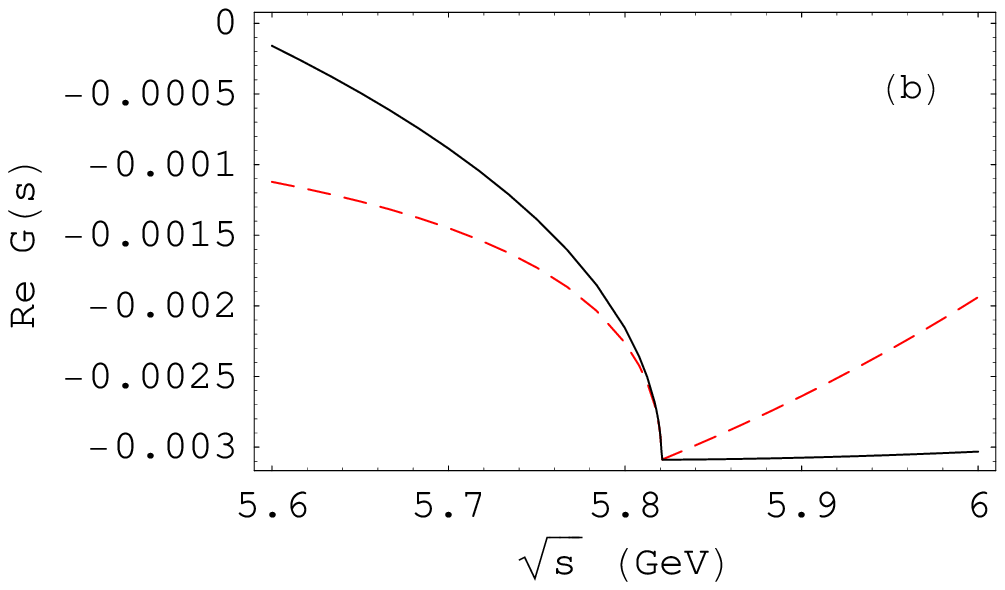}}%
\vglue -0.5cm\caption{\label{fig:match}The real parts of the loop
integrals calculated by using the cut-off method (dashed lines) and
the dispersion relation method (solid lines). (a) $D^*K$ loop, (b)
$B^*{\bar K}$ loop.}
\end{center}
\end{figure}

In ChUA, the unitary scattering amplitude can be expressed by the
algebraic Bethe-Salpeter equation \cite{oo97}. The full unitary
amplitude for the $S$ wave scattering of vector and light
pseudoscalar mesons can be written as \cite{kl04,ro05}
\begin{equation}
\label{eq:tvp}
T^{I}(s)=-[1+V^{I}(s)G(s)(1+\frac{q_{on}^2}{3M_V^2})]^{-1}V^{I}(s),
\end{equation}
where the polarization vectors are dropped because they are
irrelevant to the pole searching. In the equation, $M_V$ is the mass
of the vector meson in the meson loop and $q_{on}$ represents the
on-shell three-momentum in the center of mass frame. $V^I(s)$ is in
the matrix form with its elements being the $S$ wave projections of
$V^I_{ij}(s,t,u)$. The non-zero element of the diagonal matrix
$G(s)$ is the two-meson loop integral
\begin{equation}
G_{ii}(s)=i\int\frac{d^4q}{(2\pi)^4}\frac{1}{q^2-m_{\phi}^2+i
\varepsilon} \frac{1}{(p_1+p_2-q)^2-M_V^2+i\varepsilon},
\label{eq:2loop}
\end{equation}
where $m_{\phi}$ is the mass of the light pseudoscalar meson in the
loop. The loop integral is calculated in terms of the dispersion
relation with a pre-selected subtraction constant $a(\mu)$
\cite{oond}. The subtraction constant can be fixed by matching the
calculated loop integral with that calculated by using the
three-momentum cut-off method \cite{gs06}.
The matching point is taken at $M_{D^*}+m_{K}$ for the charmed
sector and $M_{B^*}+m_{K}$ for the bottom sector, respectively,
because we are interested in the energy region around the
threshold. With the same consideration shown in Ref. \cite{gs06},
the estimated three-momentum cutoff $q_{max}$ is in the region of
$0.8\pm0.2$ GeV. The corresponding values of $a(\mu)$ and
$q_{max}$ are tabulated in Table \ref{tab:a}. The resultant loop
integration curves in two different methods are plotted in Fig.
\ref{fig:match}. It is shown that they are very similar with each
other in the region around the matching point. With the estimated
$a(\mu)$ values, the unitary scattering amplitudes can be
calculated.

\begin{table}[hbt]
\caption{\label{tab:a} Corresponding $a(\mu)$ values and $q_{max}$
values with $\mu=m_{D}$ for the charmed sector and $\mu=m_{B}$ for
the bottom sector, respectively.}
\begin{center}
\begin{tabular}{cccc}
\hline\hline
$q_{max}$ (GeV) &~~ 0.6 &~~ 0.8 &~~ 1.0 \\
\hline
$a(m_D)$~~ &~~ -0.639~~ &~~ -0.714~~ &~~ -0.752 \\
$a(m_B)$~~ &~~~ -0.0764~ &~~ -0.101~ &~~ -0.113 \\
\hline\hline
\end{tabular}
\end{center}
\end{table}

The poles of the full scattering amplitudes in the $I=0,~S=1$
channel in both the charmed sector and bottom sector are searched
first. It is shown that on the first Riemann sheet of the energy
plane, there is a pole located in the real axis below the lowest
threshold of the coupled channels in either the charmed sector or
the bottom sector. The resultant pole positions with different
$a(\mu)$, which correspond to $q_{max}=0.6$, 0.8 and 1.0 GeV , are
tabulated in Table \ref{tab:i0}, respectively.
\begin{table}[hbt]
\caption{\label{tab:i0} Poles in the $(I,~S)=(0,~1)$ case.}
\begin{center}
\begin{tabular}{cccc}
\hline\hline
$q_{max}$ (GeV)~~~~ &~~ 0.6 &~~ 0.8 &~~ 1.0 \\
\hline
$D_{s1}$ (GeV)~~~~  &~~ 2.472 ~~&~~ 2.459 ~~&~~ 2.452 \\
$B_{s1}$ (GeV)~~~~  &~~ 5.785 ~~&~~ 5.775 ~~&~~ 5.771 \\
\hline\hline
\end{tabular}
\end{center}
\end{table}
These poles are apparently associated with the $D^*K$ bound state
and the $B^*{\bar K}$ bound state, respectively. More
specifically, when $a(m_D)=-0.714$, corresponding to $q_{max}=0.8$
GeV, the mass of the $D^*K$ state ($D_{s1}$) is about 2459 MeV,
which is quite consistent with the measured value of the
$D_{s1}(2460)$ state \cite{pdg06}
\begin{equation}
M_{D_{s1}(2460)}=2458.9\pm0.9 ~\text{MeV}.
\end{equation}
Taking into account the uncertainty of the subtraction constant, the
resultant masses of the $D_{s1}$ state and the undiscovered
$B^*{\bar K}$ bound state, namely $B_{s1}$, are
\begin{eqnarray}
M_{D_{s1}}&\!=&\! 2.462\pm0.010 ~\text{GeV}, \nonumber\\
M_{B_{s1}}&\!=&\! 5.778\pm0.007 ~\text{GeV}.
\end{eqnarray}
The predicted mass of the $B^*{\bar K}$ bound state is consistent
with the simple estimate of 5778 MeV in Ref. \cite{ro06}.

In order to show the effect of the coupled $D_s\eta$ ($B_s\eta$)
channel explicitly, we also present the single $D^*K$ ($B^*{\bar
K}$) channel result in Table \ref{tab:onec}. It is shown that the
single channel results are slightly larger than the coupled channel
results. It implies that $D_{s1}(2460)$ ($B_{s1}$) can be regarded
as a bound state of $D^*K$ ($B^*\bar{K}$) with a tiny component of
$D_s\eta$ ($B_s\eta$).
\begin{table}[hbt]
\caption{\label{tab:onec} Poles in the $(I,~S)=(0,~1)$ case in the
single channel approximation.}
\begin{center}
\begin{tabular}{cccc}
\hline\hline
$q_{max}$ (GeV)~~~~ &~~ 0.6 &~~ 0.8 &~~ 1.0 \\
\hline
$D_{s1}$ (GeV)~~~~  &~~ 2.478 ~~&~~ 2.467 ~~&~~ 2.462 \\
$B_{s1}$ (GeV)~~~~  &~~ 5.786 ~~&~~ 5.779 ~~&~~ 5.775 \\
\hline\hline
\end{tabular}
\end{center}
\end{table}

In the $I=1/2,~S=0$ case, the poles are located on nonphysical
Riemann sheets. Usually, for a certain energy if Im$p_{cm}$ is
negative in all the opened channels, the pole obtained would
correspond more closely with the physical one. There are two poles
in the charmed (bottom) sector. The width of the lower pole is broad
and the width of the higher one is relatively narrow. We tabulated
these results in Table \ref{tab:i1/2}.
\begin{table}[hbt]
\caption{\label{tab:i1/2} Poles in the $(I,~S)=(\frac{1}{2},~0)$
case.}
\begin{center}
\begin{tabular}{cccc}
\hline\hline
$q_{max}$ (GeV) & 0.6 & 0.8 & 1.0 \\
\hline
\multirow{2}{*}{$D_{1}$ (GeV)}  &~~ $2.245-i0.106$ ~~&~~ $2.239-i0.094$ ~~&~~ $2.236-i0.088$ \\
                                &~~ $2.599-i0.043$ ~~&~~ $2.585-i0.044$ ~~&~~ $2.578-i0.043$ \\
\hline
\multirow{2}{*}{$B_{1}$ (GeV)}  &~~ $5.586-i0.118$ ~~&~~ $5.579-i0.108$ ~~&~~ $5.576-i0.103$ \\
                                &~~ $5.884-i0.027$ ~~&~~ $5.875-i0.025$ ~~&~~ $5.870-i0.023$ \\
\hline\hline
\end{tabular}
\end{center}
\end{table}
In the charmed (bottom) sector, the lower pole is located on the
second Riemann sheet (Im$p_{cm1}<0$, Im$p_{cm2}>0$, Im$p_{cm3}>0$,
where $p_{cmi}$ is the momentum of one of the interacting mesons in
the center of mass frame in the $i$-th channel) and should be
associated with the $D^*\pi$ ($B^*\pi$) resonance. This state should
easily decay into $D^*\pi$ ($B^*\pi$). The higher pole in the
charmed (bottom) sector is found on the third Riemann sheet
(Im$p_{cm1}<0$, Im$p_{cm2}<0$, Im$p_{cm3}>0$) and should be
associated with an ``unstable" $D_s^*{\bar K}$ ($B_s^*K$) bound
state due to its narrow width. It should be mentioned that the pole
structures of $1^+$ states here are very similar to those of $0^+$
states \cite{gs06}, but are different from that of the $f_0(980)$
state where only one pole located on the second Riemann sheet and
one shadow pole on the third Riemann sheet \cite{zb93}. The origin
of the difference comes from the fact that there are two coupled
channels, i.e. the $\pi\pi$ and $K{\bar K}$ channels, in the
$f_0(980)$ state case, while there are three coupled channels in the
$1^+$ state case.

\begin{figure}[hbt]
\begin{center}\vspace*{0.5cm}
{\epsfysize=4.5cm \epsffile{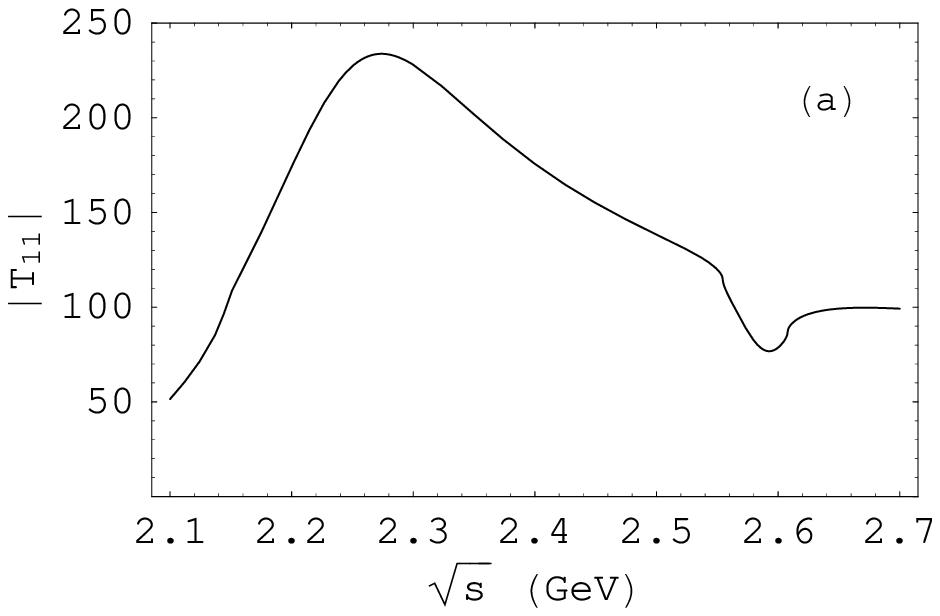}}%
{\epsfysize=4.5cm \epsffile{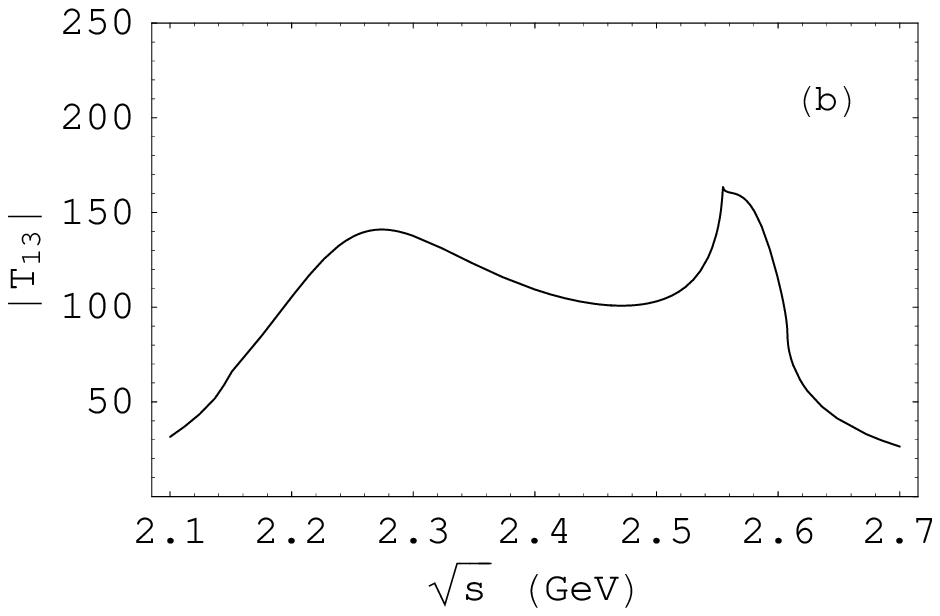}}%
\vglue -0.5cm\caption{\label{fig:T11_c}The absolute values of
unitary scattering amplitudes in the coupled-channel case for the
$D^*\pi\to D^*\pi$ and $D^*_s{\bar K}\to D^*\pi$ processes in the
$I=1/2$ channel. (a) $D^*\pi\to D^*\pi$, (b) $D^*_s{\bar K}\to
D^*\pi$.}
\end{center}
\end{figure}

The fact that two poles in the different Riemann sheets should be
associated with two different $1^+$ states in the $I=1/2$ channel
can be confirmed by checking the curve structure of the absolute
values of the unitary scattering amplitudes in the coupled channel
case, because such curve structure is closely related to the
structure in the corresponding invariant mass spectrum. The absolute
values of the unitary scattering amplitudes in the coupled-channel
case for the $D^*\pi\to D^*\pi$ and $D^*_s{\bar K}\to D^*\pi$
processes in the $I=1/2$ channel are plotted in Figs.
\ref{fig:T11_c} (a) and (b), respectively. It is shown that there
are one broad peak and one narrow dip in the $D^*\pi\to D^*\pi$
scattering amplitude and two peaks in the $D^*_s{\bar K}\to D^*\pi$
case. In the bottom sector, the curve structure of scattering
amplitudes is the same and will not be demonstrated here for
simplicity.

Considering the uncertainty of $a(\mu)$ in Table \ref{tab:i1/2}, we
predict the masses and the widths of the broad $D_{1}$ and $B_1$
states as
\begin{eqnarray}
M_{D_{1}}&\!=&\! 2.240\pm0.005 ~\text{GeV},\quad \Gamma_{D_{1}}=0.194\pm0.019 ~\text{GeV}, \nonumber\\
M_{B_{1}}&\!=&\! 5.581\pm0.005 ~\text{GeV},\quad \Gamma_{B_{1}}=0.220\pm0.015 ~\text{GeV},%
\end{eqnarray}
respectively, and the masses and the widths of the narrow $D_{1}$
and $B_{1}$ states as
\begin{eqnarray}
M_{D_{1}}&\!=&\! 2.588\pm0.010 ~\text{GeV},\quad \Gamma_{D_{1}}=0.087\pm0.001 ~\text{GeV}, \nonumber\\
M_{B_{1}}&\!=&\! 5.877\pm0.007 ~\text{GeV},\quad \Gamma_{B_{1}}=0.050\pm0.003 ~\text{GeV},%
\end{eqnarray}
respectively.

These results mean that if we believe that the $D_{s0}^*(2317)$ and
$D_{s1}^*(2460)$ states in the $(I,~S)=(0,~1)$ channel are really
dominated by the molecular state structure, the predicted two $1^+$
states in the $(I,~S)=(1/2,~0)$ channel should also exist. As
mentioned in the Introduction, the experimentally established $1^+$
states $D_1(2420)$ and $D_1(2430)$ are compatible with the
conventional $c{\bar q}$ interpretation. It implies that there are
no experimentally established states as the candidates for the
predicted $(I,~S)=(1/2,~0)$ states. Why these two states have not
been observed are complex. At the present time, it is not at all
clear about what to expect with regards to production of these
states. In fact, finding a new state does not only depend on its
high production rate, but also relate to, in a large extent, the
data measurement and analysis, which are usually affected by many
factors, for instance the data statistics, the background, the width
of the state, the complexity of the spectrum structure in the
vicinity of the state, and the suitable channels for producing and
detecting such a state, and etc.. For instance, one of the possible
reasons which makes the observation of the predicted states
difficult is that the production of such states would be suppressed
with respect to conventional ones. This is because: (1) The
predicted two $D_1$ states are quasi-bound states of other two
mesons. From the viewpoint of quark degrees of freedom, there should
be at least four quarks in the Fock space. Therefore, the production
of such states would be suppressed in comparison with producing a
$c{\bar q}$ state due to the necessary creation of an additional
quark anti-quark pair. (2) The widths of predicted $D_1$ states are
comparable with those of the corresponding conventional $c\bar{q}$
states, so that the couplings between these states to the $D^*\pi$
state would be similar with the $c\bar{q}$ states. In other words,
the signals of the predicted $D_1$ states would be suppressed
compared with those of the conventional $c{\bar q}$ $D_1$ states in
the $D^*\pi$ spectrum, and a definite observation of such states
becomes difficult. It seems that an even larger data set with higher
statistics and further careful data analysis, as well as theoretical
study are necessary.

Moreover, it is interesting to compare our predicted states with the
quark model predicted conventional $q{\bar q}$ states shown in Table
\ref{tab:qm} \cite{gi85,gk91,pe01}. We find that the lowest masses
of the conventional $Q{\bar s}$ states with $J^P=1^+$ are larger
than those of our quasibound $D_{s1}$ and $B_{s1}$ states,
respectively, and the lowest masses of the conventional $Q{\bar q}$
($q=u,d$) states with $J^P=1^+$ are sited between the masses of our
lower and higher predicted states. Hence, if one can find a state
with a mass much lower than the quark model prediction, it might be
the lower state in our prediction.
\begin{table}[hbt]
\caption{\label{tab:qm} Masses of $1^1P_1$ and $1^3P_1$ heavy-light
mesons in quark models.}
\begin{center}
\begin{tabular}{cccc}
\hline\hline
 & Ref. \cite{gi85} & Ref. \cite{gk91} & Ref. \cite{pe01} \\\hline
$D_1(1^1P_1)$ & 2.44 & 2.46 & 2.490 \\
$D_1(1^3P_1)$ & 2.49 & 2.47 & 2.417 \\
$D_{s1}(1^1P_1)$ & 2.53 & 2.55 & 2.605 \\
$D_{s1}(1^3P_1)$ & 2.57 & 2.55 & 2.535 \\
$B_1(1^1P_1)$ & & 5.78 & 5.742 \\
$B_1(1^3P_1)$ & & 5.78 & 5.700 \\
$B_{s1}(1^1P_1)$ & & 5.86 & 5.842 \\
$B_{s1}(1^3P_1)$ & & 5.86 & 5.805 \\
\hline\hline
\end{tabular}
\end{center}
\end{table}

\section{Coupling constants and decay widths}

The nature of dynamically generated states can be further studied by
calculating the coupling constants between these states and the
particles in the coupled channels. The coupling constants are
related to the Laurent expansions of unitary scattering amplitudes
around the pole \cite{ol05}
\begin{equation}
T_{ij}=\frac{g_ig_j}{s-s_{pole}}+\gamma_0+\gamma_1(s-s_{pole})+\cdots,
\end{equation}
where $g_i$ and $g_j$ are the coupling constants of the generated
state to the $i$-th and $j$-th channels. The product $g_ig_j$ can
be obtained by calculating the residue of the unitary scattering
amplitude at the pole \cite{oond}
\begin{equation}
g_ig_j=\lim_{s\to s_{pole}}(s-s_{pole})T_{ij}.
\end{equation}

As a typical example, we calculate the coupling constants for the
$D_{s1}(2460)$ state with a subtraction constant which corresponds
to $q_{max}=0.8$ GeV, because under this condition the empirical
mass of $D_{s1}(2460)$ can be excellently reproduced. The
resultant coupling constants are tabulated in Table \ref{tab:ccs1}
for the states with $I=0$ and Table \ref{tab:ccs0} for the states
with $I=1/2$.

Comparing the coupling constants in Tables \ref{tab:ccs1} and
\ref{tab:ccs0} with theose of the generated scalar states in Tables
VI and VII in Ref. \cite{gs06}, one finds that the values of
corresponding coupling constants are close. This manifests the heavy
flavor spin symmetry \cite{ne94} respected in the leading order
heavy chiral Lagrangian \cite{bd92}.
\begin{table}[hbt]
\caption{\label{tab:ccs1} Coupling constants of the generated
$D_{s1}$ and $B_{s1}$ states to relevant coupled channels. In this
case, $g_1$ and $g_2$ are real. All units are in GeV.}
\begin{center}
\begin{tabular}{cccc}
\hline\hline $~$ & Masses &~~~$|g_1|$~~~&~~~$|g_2|$\\
\hline $D_{s1}$ ~~~~&~~2.459~~&~~10.762~~&~~6.170  \\ \hline
$B_{s1}$ ~~~~&~~5.775~~&~~23.572~~&~~13.326  \\
\hline\hline
\end{tabular}
\end{center}
\end{table}
\begin{table}[hbt]
\caption{\label{tab:ccs0} Coupling constants of the generated
$D_{1}$ and $B_{1}$ states to relevant coupled channels. All units
are in GeV.}
\begin{center}
\begin{tabular}{cccccccc}
\hline\hline
 $~$ &  Poles  & $g_1$ & $|g_1|$ & $g_2$ & $|g_2|$ & $g_3$ & $|g_3|$\\
\hline $D_{1}$ & $2.239-i0.094$ & $8.157+i5.135$ & 9.639 & $-0.202+i0.059$%
                            & 0.211 & $4.919+i3.053$ & 5.790 \\
       $D_{1}$ & $2.585-i0.044$ & $0.145+i3.306$ & 3.309 & $-6.893-i2.237$%
                            & 7.247 & $-11.060+i1.165$ & 11.121 \\
\hline $B_{1}$ & $5.579-i0.108$ & $21.439+i11.861$ & 24.502 & $-2.222-i0.752$%
                            & 2.346 & $13.517+i6.906$ & 15.179 \\
       $B_{1}$ & $5.875-i0.025$ & $0.295+i6.619$ & 6.626 & $-14.553-i4.892$%
                            & 15.353 & $-24.759-i0.874$ & 24.775 \\
\hline\hline
\end{tabular}
\end{center}
\end{table}
In Table \ref{tab:ccs1}, the data also show that $g_1$ is larger
than $g_2$ in the charmed (bottom) sector. This indicates that the
coupling between the $D_{s1}$ ($B_{s1}$) state and the states in the
$D^*K$ ($B^*{\bar K}$) channel is larger than that between the
$D_{s1}$ ($B_{s1}$) state and the states in the $D^*_s\eta$
($B^*_s\eta$) channel. It also reflects the fact that the generated
state is a $D^*K$ ($B^*{\bar K}$) bound state. In Table
\ref{tab:ccs0}, the largest coupling constant $|g_1|$ for the lower
state is associated with the $D^*\pi$ ($B^*\pi$) channel and the
largest coupling constant $|g_3|$ for the higher state is connected
with the $D_s^*{\bar K}$ ($B_s^*K$) channel. This is consistent with
our finding in the pole analysis, namely the lower state is a
$D^*\pi$ ($B^*\pi$) resonance and the higher state associates with a
$D^*_s{\bar K}$ ($B^*_sK$) quasi-bound state .

The coupling constants also show that the largest component of the
lower $D_{1}$ state is $D^*\pi$ whose quark contents are $c{\bar
n}n{\bar n}$, where $n$ denotes the $u$ or $d$ quark, and the
largest component of the higher $D_{1}$ state is $D_s^*\bar{K}$
whose quark contents are $c{\bar s}s{\bar s}$. On the other hand,
the $D_{s1}(2460)$ state, in principle, is a $D^*K$ bound state, and
consequently, the dominant quark contents of this state are $c{\bar
n}n{\bar s}$. Thus, from the quark contents of these states, one can
expect that the mass of $D_{s1}$ should be a value between the
masses of the two $D_1$ states. The same qualitative statement can
be applied to the bottom sector.

Furthermore, the dynamic nature of a state can be characterized by
its decay properties. Since the quark contents of predicted
molecular states are different from those of conventional $q{\bar
q}$ states, their decay properties are expected to be different.

\begin{figure}[htb]
\begin{center}\vspace*{0.5cm}
{\epsfysize=3.5cm \epsffile{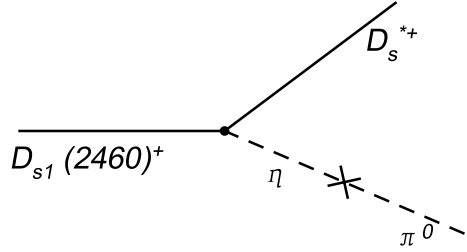}}%
\vglue -0.5cm\caption{\label{fig:isv}Isospin symmetry violating
decay process $D_{s1}(2460)^+\to D^{*+}\pi^0$ via $\pi^0$-$\eta$
mixing.}
\end{center}
\end{figure}

In order to estimate the order of magnitude of the decay width of
$D_{s1}(2460)$, we calculate the decay widths of the isospin
symmetry violating decay processes $D_{s1}(2460)^+\to D^{*+}\pi^0$
and $B_{s1}(5775)^0\to B^{*0}\pi^0$ can be estimated through
$\pi^0$-$\eta$ mixing \cite{cw94} shown in Fig. \ref{fig:isv}.
By using the formula
\begin{equation}
\Gamma=\frac{p_{cm}}{8\pi M^2}
\overline{\sum_{\lambda}}\sum_{\lambda'}
|\frac{g_2t_{\pi\eta}\epsilon^{\lambda}(p)\cdot\epsilon^{\lambda'*}(p_1)}{m_{\pi^0}^2-m_{\eta}^2}|^2,
\end{equation}
where $M$ is the mass of the initial meson, $p_{cm}$ denotes the
three-momentum in the center of mass frame,
$\epsilon^{\lambda}(p)$ and $\epsilon^{\lambda'}(p_1)$ represent
the polarization vectors of the $D_{s1}(2460)$ ($B_{s1}(5775)$)
and $D^*$ ($B^*$) states, respectively, and
$\overline{\sum}_{\lambda}\sum_{\lambda'}$ describes the sum over
the final states and the average over the initial states. By using
the $\pi^0$-$\eta$ mixing amplitude $t_{\pi\eta}=-0.003$ GeV
obtained from the Dashen's theorem \cite{dash}, we get the decay
widths
\begin{eqnarray}
\Gamma(D_{s1}(2460)^{+}\to D_s^{*+}\pi^0)&\!=&\! 11.41~\text{keV},\nonumber\\
\Gamma(B_{s1}(5775)^{0}\to B_s^{*0}\pi^0)&\!=&\! 10.36~\text{keV}.
\end{eqnarray}
It should be mentioned that, the explicit isospin breaking term is
not the only source of isospin breaking, the mass difference of
isospin multiplets has already generated such a breaking. Thus, the
given result can only serve an estimate of the order of magnitudes
of the widths.

The decay properties of predicted $D_1$ ($B_1$) state can also be
briefly estimated. For the higher $D_1$ ($B_1$) states, it can
strongly decay into the two opened channels: $D^*\pi$ and $D^*\eta$
($B^*\pi$ and $B^*\eta$). Although the channel with $\eta$ have much
smaller phase space than that with $\pi$, due to the relatively
larger contribution from hidden strangeness, which can be seen from
Table \ref{tab:ccs0}, the branching fraction of the $D_1$ ($B_1$)
state decaying into the final state with $\eta$ is comparable with
or even larger than that with $\pi$. The ratio of these two
branching fractions can roughly be estimated by using corresponding
coupling constants given in Table \ref{tab:ccs0}
\begin{eqnarray}
\label{eq:ratio} R_{\eta/\pi^0}(D_1)\equiv \frac{\Gamma(D_1\to
D^*\eta)}{\Gamma(D_1\to D^*\pi)}
\simeq 1.57 \nonumber\\
R_{\eta/\pi^0}(B_1)\equiv \frac{\Gamma(B_1\to
B^*\eta)}{\Gamma(B_1\to B^*\pi)} \simeq 0.50.
\end{eqnarray}
As to the conventional $D_1$ ($B_1$) state which has a mass larger
than the $D^*\eta$ ($B^*\eta$) threshold, the branching fraction of
decaying into $D^*\eta$ ($B^*\eta$) would be much smaller than that
into $D^*\pi$ ($B^*\pi$) due to both OZI suppression \cite{ozi} and
phase space suppression. Thus, it would be easy to distinguish this
state from the higher $D_1$ ($B_1$) state by measuring the ratio
defined in Eq. (\ref{eq:ratio}). For the lower $D_1$ ($B_1$) state
predicted here, it has only one opened channel $D^*\pi$ ($B^*\pi$).
So it would be difficult to distinguish the lower state from the
conventional $D_1$ ($B_1$) state with similar mass and width by
using strong decay properties. Yet, the radiative decays of such
states might be different. The concrete consequences should be
investigated in future.

\section{Conclusion}

We study the dynamically generated axial heavy mesons which have
the same quantum numbers of the conventional $c{\bar s}$ and
$c{\bar q}$ states in the $(I,S)=(0,1)$, $(I,S)=(1/2,0)$ systems
in the framework of coupled-channel HChUA. In the $(I,S)=(0,1)$
and $J^P=1^+$ system, there are two coupled channels: $D^*K$ and
$D_s^*\eta$ for the charmed sector and $B^*\bar{K}$ and
$B_s^*\eta$ for the bottom sector, respectively. In the coupled
channel calculation, the channels with a light vector meson and a
heavy pseudoscalar meson are not considered due to their less
importance.

By searching for the pole of the unitary coupled-channel scattering
amplitude on appropriate Riemann sheets of the energy plane, we find
a state with mass of $2.462\pm0.010$ GeV in the $(I,S)=(0,1)$
system. This state should be a $D^*K$ bound state with a tiny
$D_s^*\eta$ component. We would interpret it as the recently
observed charmed meson $D_{s1}(2460)$. In the same way, we predict a
$B^*{\bar K}$ bound state with mass of $5.778\pm0.007$ GeV in the
bottom sector. In the $(I,S)=(1/2,0)$, $J^P=1^+$ system, there are
three coupled channels: $D^*\pi$, $D^*\eta$ and $D_s^*{\bar K}$ in
the charmed sector and $B^*\pi$, $B^*\eta$ and $B_s^*K$ in the
bottom sector, respectively. We find two poles in the nonphysical
Riemann sheets in both the charmed and bottom sectors. In the
charmed (bottom) sector, the lower pole is located at
$(2.240\pm0.005-i0.097\pm0.009)$ GeV
($(5.581\pm0.005-i0.110\pm0.007)$ GeV). The state associated with
this pole will strongly decay to $D^*\pi$ ($B^*\pi$) and have the
largest coupling with the decayed particles. Therefore, this pole
should be associated with the $D^*\pi$ ($B^*\pi$) resonance. The
higher pole is positioned at $(2.588\pm0.010-i0.043\pm0.001)$ GeV
($(5.877\pm0.007-i0.025\pm0.002)$ GeV). The state associated with
this pole has two decay channels $D^*\pi$ and $D^*\eta$ ($B^*\pi$
and $B^*\eta$) and has the largest coupling with decayed particles
in the $D^*_s{\bar K}$ ($B^*_sK$) channel. Thus, this pole should be
associated with a quasi-bound state of $D^*_s{\bar K}$ ($B^*_sK$).
If one believes that the corresponding states in the $(I,~S)=(0,~1)$
and $(I,~S)=(1/2,~0)$ systems have similar $S$ wave molecular state
structures, the two predicted states should exist. The estimated
order of magnitudes for the widths of the $D_{s1}(2460)^{+}$ and
$B_{s1}(5775)^{0}$ states is about 10 keV. The decay properties of
predicted $D_1$ ($B_1$) states are also briefly discussed. Study the
channels where the final states include $D^*\eta$ and $D^*\pi$
($B^*\eta$ and $B^*\pi$) would be helpful to find predicted $D_1$
($B_1$) states.

\begin{acknowledgments}
We sincerely thank Prof. B.S. Zou for valuable discussion. We thank
G. Rupp for helpful comments on their work and pointing out a
typographic error in the previous version of this paper. This work
is partially supported by the NSFC grant Nos. 90103020, 10475089,
10435080, 10447130, CAS Knowledge Innovation Key-Project grant No.
KJCX2SWN02 and Key Knowledge Innovation Project of IHEP, CAS (U529).
\end{acknowledgments}


\end{document}